\documentclass[twocolumn,showkeys,aps,prb,showpacs]{revtex4-1}
\usepackage{graphicx}
\usepackage{amsmath}
\usepackage[CJKbookmarks,dvipdfm,colorlinks,linkcolor=blue,citecolor=blue]{hyperref}

\begin{document}

\title{Born effective charge removed anomalous temperature dependence of lattice thermal conductivity in monolayer GeC}

\author{San-Dong Guo}
\affiliation{School of Physics, China University of Mining and
Technology, Xuzhou 221116, Jiangsu, China}

\begin{abstract}
Due to potential applications in nano- and opto-electronics,  two-dimensional (2D) materials have attracted tremendous interest. Their thermal transport properties are closely related to the performance of  2D materials-based devices.
Here, the phonon transports of  monolayer GeC with a perfect planar
hexagonal honeycomb structure  are investigated by solving  the linearized phonon Boltzmann equation within the single-mode relaxation time approximation (RTA). Without inclusion of  Born effective charges ($Z^*$) and dielectric constants  ($\varepsilon$),  the lattice  thermal conductivity ($\kappa_L$) almost decreases linearly above 350 K, deviating  from the usual $\kappa_L$$\sim$$1/T$ law. The underlying mechanism  is because the  contribution  to $\kappa_L$ from high-frequency optical phonon modes increases with increasing temperature, and the contribution exceeds one from acoustic branches at high temperature.
These can be understood by huge phonon band gap caused by large difference in atom mass between Ge and C atoms, which produces important effects on scattering process  involving high-frequency optical phonon.
When considering $Z^*$ and $\varepsilon$, the phonon group velocities and phonon lifetimes of high-frequency optical phonon modes are obviously reduced with respect to ones without $Z^*$ and $\varepsilon$. The reduced group velocities and phonon lifetimes give rise to small contribution  to $\kappa_L$ from high-frequency optical phonon modes, which produces the the traditional $\kappa_L$$\sim$$1/T$ relation in monolayer GeC. Calculated results show that
the isotope scattering can also reduce anomalous temperature dependence of $\kappa_L$ in monolayer GeC.  Our works highlight the importance of $Z^*$ and $\varepsilon$ to investigate phonon transports of monolayer GeC, and  motivate further theoretical or experimental efforts to investigate thermal transports of other 2D materials.

\end{abstract}

\pacs{72.15.Jf, 71.20.-b, 71.70.Ej, 79.10.-n ~~~~~~~~~~~~~~~~~~~~~~~~~~~~~~~~~~~Email:sandongyuwang@163.com}

\maketitle

\section{Introduction}
Since the successful synthesis of graphene\cite{q1}, 2D materials  have attracted increasing attention due to potential  applications in electronics, spintronics and optoelectronics\cite{q2,q3,q4,q5,q6,q7,q8,q9}.  Thermal management is a significant factor for these applications\cite{q10}. To efficiently dissipate heat  in electronic devices, a high thermal conductivity is required, while a low lattice thermal conductivity is beneficial to thermoelectric applications,  to achieve good thermoelectric performance\cite{q11,q12}.
 Diverse anisotropy of phonon transport in  group IV-VI monolayer is predicted by  solving the Boltzmann transport equation\cite{q15}.
Phonon transport properties of group-IV and -VA element  monolayers have been performed theoretically\cite{q16,q17,q17-1}.
 The $\kappa_L$ of  transition metal dichalcogenide (TMD) and Janus TMD monolayers have been systematically studied by  phonon Boltzmann
transport equation approach\cite{p5,p6}.
Strain effects on thermal transports  of Sb monolayer\cite{q18},  group-IV monolayers\cite{q19}  and 2D penta-structures materials\cite{q20} have also been studied, showing  diverse strain dependence, such as  monotonously increasing or decreasing  and   up-and-down  behaviors with increasing tensile strain.
\begin{figure}
  \includegraphics[width=5.5cm]{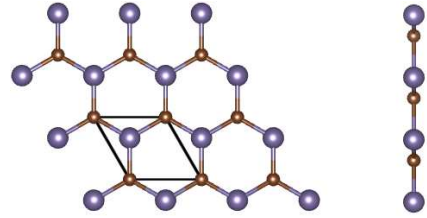}
  \caption{(Color online) The top (Left) and side (Right) view of monolayer GeC, and the frame surrounded by a black box is  unit cell. The large  and small balls represent Ge and C atoms, respectively.}\label{st}
\end{figure}

For most  bulk and 2D materials, the temperature-dependent $\kappa_L$  follows the relation of $\kappa_L$$\sim$$1/T$.
Recently, anomalously temperature-dependent $\kappa_L$ of monolayer ZnO and GaN is predicted by a first-principles study\cite{q21,q22}, which is due to the huge phonon band gap in their phonon dispersions.
The predicted room-temperature $\kappa_L$  of monolayer ZnO and GaN is 4.5  $\mathrm{W m^{-1} K^{-1}}$ with the thickness of 3.04 $\mathrm{{\AA}}$ and  14.93 $\mathrm{W m^{-1} K^{-1}}$ with the thickness of 3.74 $\mathrm{{\AA}}$, respectively.
The SiC monolayer has the same perfect planar hexagonal honeycomb structure with ZnO and GaN,  but the $\kappa_L$ of SiC monolayer follows the conventional $1/T$ law\cite{q22-1}, which may be due to small phonon band gap. For GeC monolayer with the same structure, a large phonon band gap can be observed  due to a large difference in atom mass between Ge and C atoms\cite{q22-2}. The similar anomalous temperature dependence of $\kappa_L$  may also exist in monolayer GeC. In this work,   based on first-principles calculations, the phonon transport properties of
monolayer GeC are investigated  by solving the linearized phonon Boltzmann equation.
 When neglecting $Z^*$ and $\varepsilon$,  the $\kappa_L$  deviates from the usual $\kappa_L$$\sim$$1/T$ law. The $\kappa_L$
above 200 K is much higher than the expected $\kappa_L$ predicted from  the general  $\kappa_L$$\sim$$1/T$ law. The large deviation  stems from the high-frequency optical phonon modes, whose
contribution to $\kappa_L$ increases with increasing
temperature,  and eventually dominates  $\kappa_L$.
 With inclusion of $Z^*$ and $\varepsilon$, the phonon group velocities and phonon lifetimes of high-frequency optical phonon modes are obviously reduced, which  gives rise to small contribution  to $\kappa_L$ from high-frequency optical phonon modes,  producing  the the traditional $\kappa_L$$\sim$$1/T$ relation in monolayer GeC. It is found  that the isotope scattering can also reduce anomalous temperature dependence of $\kappa_L$ in monolayer GeC.

The rest of the paper is organized as follows. In the next
section, we shall give our computational details about phonon transport. In the third section, we shall present phonon transport of monolayer GeC. Finally, we shall give our discussion and conclusions in the fourth section.

\begin{figure}
  \includegraphics[width=8cm]{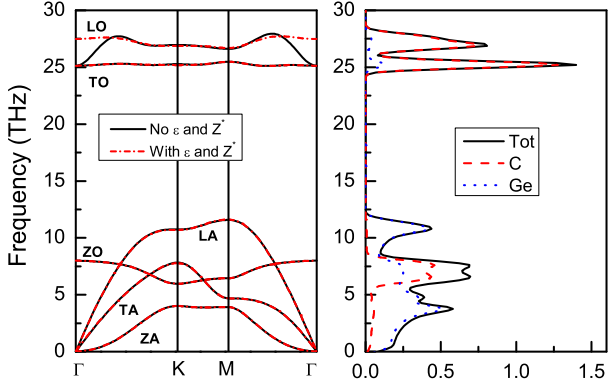}
  \caption{(Color online) Phonon band structures of monolayer GeC with or without  $\varepsilon$ and $Z^*$, along with the total and partial DOS without $\varepsilon$ and $Z^*$ . }\label{ph}
\end{figure}

\section{Computational detail}
All first-principles calculations are carried out based
on the density functional theory (DFT) using the projected augmented wave (PAW) method, and the generalized gradient approximation of the Perdew-Burke-Ernzerhof (GGA-PBE) is adopted as  exchange-correlation energy functional,
as implemented in the  Vienna ab initio simulation package (VASP)\cite{pv1,pv2,pbe,pv3}.
A plane-wave basis set is employed with
kinetic energy cutoff of 700 eV, and the $2s2p$ ($4s4p$) orbitals of C (Ge) atoms are treated as valance ones. To avoid spurious interaction,
the unit cell  of monolayer GeC  is built with the vacuum region of  18 $\mathrm{{\AA}}$ along the
out-of-plane direction. The energy convergence threshold is set as  $10^{-8}$ eV.
\begin{figure}
  \includegraphics[width=8cm]{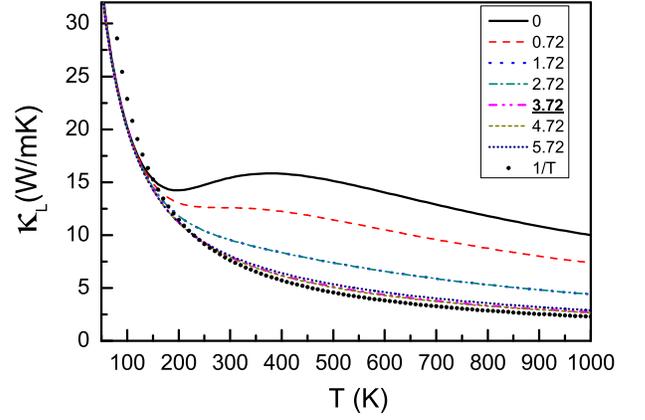}
  \caption{(Color online)  The $\kappa_L$ of  monolayer GeC  with or without  $\varepsilon$ and $Z^*$ as a function of temperature (3.72 and 0); The $\kappa_L$$\sim$$1/T$ relation is plotted for comparison;  The $\kappa_L$ with $|Z^*|$ along $xx$ and $yy$ directions artificially changing from 0.72 to 5.72.
   }\label{kl}
\end{figure}

The $\kappa_L$ of  monolayer GeC   is calculated by solving linearized phonon Boltzmann equation with the single mode RTA,   as implemented in the Phono3py code\cite{pv4}.
The $\kappa_L$ can be expressed as:
\begin{equation}\label{eq0}
    \kappa=\frac{1}{NV_0}\sum_\lambda \kappa_\lambda=\frac{1}{NV_0}\sum_\lambda C_\lambda \nu_\lambda \otimes \nu_\lambda \tau_\lambda
\end{equation}
where $\lambda$, $N$ and  $V_0$ are  phonon mode, the total number of q points sampling Brillouin zone (BZ) and  the volume of a unit cell, and  $C_\lambda$,  $ \nu_\lambda$, $\tau_\lambda$   is the specific heat,  phonon velocity,  phonon lifetime.
The phonon lifetime $\tau_\lambda$ can be attained  by  phonon linewidth $2\Gamma_\lambda(\omega_\lambda)$ of the phonon mode
$\lambda$:
\begin{equation}\label{eq0}
    \tau_\lambda=\frac{1}{2\Gamma_\lambda(\omega_\lambda)}
\end{equation}
The $\Gamma_\lambda(\omega)$  takes the form analogous to the Fermi golden rule:
\begin{equation}
\begin{split}
   \Gamma_\lambda(\omega)=\frac{18\pi}{\hbar^2}\sum_{\lambda^{'}\lambda^{''}}|\Phi_{-\lambda\lambda^{'}\lambda^{''}}|^2
   [(f_\lambda^{'}+f_\lambda^{''}+1)\delta(\omega
    -\omega_\lambda^{'}-\\\omega_\lambda^{''})
   +(f_\lambda^{'}-f_\lambda^{''})[\delta(\omega
    +\omega_\lambda^{'}-\omega_\lambda^{''})-\delta(\omega
    -\omega_\lambda^{'}+\omega_\lambda^{''})]]
\end{split}
\end{equation}
in which $f_\lambda$  and $\Phi_{-\lambda\lambda^{'}\lambda^{''}}$ are the phonon equilibrium occupancy and the strength of interaction among the three phonons $\lambda$, $\lambda^{'}$, and $\lambda^{''}$ involved in the scattering.

The interatomic force constants (IFCs) are calculated by
the finite displacement method.
 The second-order harmonic (third-order anharmonic) IFCs
are calculated using a 5 $\times$ 5 $\times$ 1 (4 $\times$ 4 $\times$ 1)  supercell  containing
50 (32) atoms with k-point meshes of 4 $\times$ 4 $\times$ 1. Using the harmonic IFCs, phonon dispersion of monolayer GeC can be attained, using Phonopy package\cite{pv5}.  To compute lattice thermal conductivities, the
reciprocal spaces of the primitive cells  are sampled using the 100 $\times$ 100 $\times$ 1 meshes.
For 2D material, the calculated  lattice  thermal conductivity  depends on the length of unit cell used in the calculations along $z$ direction\cite{2dl}.  The lattice  thermal conductivity should be normalized by multiplying $Lz/d$, in which  $Lz$ is the length of unit cell along $z$ direction  and $d$ is the thickness of 2D material, but the $d$  is not well defined like graphene.   In this work, the length of unit cell (18 $\mathrm{{\AA}}$) along z direction is used as the thickness of  monolayer GeC. To make a fair comparison between various 2D monolayers, the thermal sheet conductance can be used, defined as $\kappa_L$ $\times$ $d$.

\begin{figure}
  \includegraphics[width=8cm]{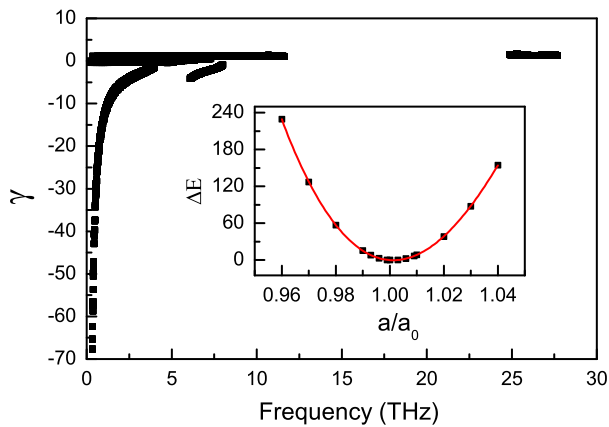}
  \caption{(Color online) The mode level Gr$\mathrm{\ddot{u}}$neisen parameters  of  monolayer GeC in the first BZ; (Inset) the potential energy well $\Delta E$=$E_{a/a_0}$-$E_1$ as a function of $a/a_0$.}\label{gru}
\end{figure}

\section{MAIN CALCULATED RESULTS AND ANALYSIS}
Monolayer GeC prefers   a perfect planar hexagonal honeycomb structure, and similar monolayer structure can be found
in graphene,  ZnO,  GaN and SiC\cite{q21,q22,q22-2}.  The monolayer
GeC  can be built by replacing one atom in the
unit cell of graphene with Ge atom, and  the space symmetry group is $P\bar{6}M2$, being
lower than that of graphene ($P6/MMM$).  \autoref{st}  shows the schematic crystal structure of monolayer GeC, and the optimized lattice parameter within GGA-PBE is 3.26 $\mathrm{{\AA}}$.  Firstly, the elastic properties of monolayer GeC are studied, and  two independent elastic
constants $C_{11}$ (=$C_{22}$) and $C_{12}$ due to $D_{3h}$ symmetry can be calculated, and the $C_{66}$=($C_{11}$-$C_{12}$)/2.
For $C_{11}$, $C_{12}$ and $C_{66}$, the calculated value is 159.42 $\mathrm{Nm^{-1}}$,  51.62  $\mathrm{Nm^{-1}}$ and 53.90 $\mathrm{Nm^{-1}}$, respectively.
These $C_{ij}$ satisfy the  Born  criteria of mechanical stability.
Based on calculated  $C_{ij}$, the 2D Young¡¯s moduli $Y^{2D}$   and shear modulus $G^{2D}$ of monolayer GeC\cite{ela} are 142.71  $\mathrm{Nm^{-1}}$ and 53.90  $\mathrm{Nm^{-1}}$, which are  lower than ones of  graphene and SiC monolayer\cite{ela,q22-1}. The GeC monolayer is more flexible  than   graphene and SiC monolayer  due to  smaller $Y^{2D}$.

\begin{table}[!htb]
\centering \caption{The $Z^*$ of C and Ge atoms and $\varepsilon$ of monolayer GeC. Except for $xx$, $yy$ and $zz$ directions,  the $Z^*$ and $\varepsilon$ along other directions are zero.}\label{tab1}
  \begin{tabular*}{0.48\textwidth}{@{\extracolsep{\fill}}cccc}
  \hline\hline
Direction & $Z^*$(C) &$Z^*$(Ge)& $\varepsilon$ \\\hline\hline
$xx$ &-3.72&3.72& 2.90\\
$yy$ &-3.72&3.72& 2.90\\
$zz$ &-0.25&0.25&1.18\\\hline\hline
\end{tabular*}
\end{table}

\begin{figure}
  \includegraphics[width=8cm]{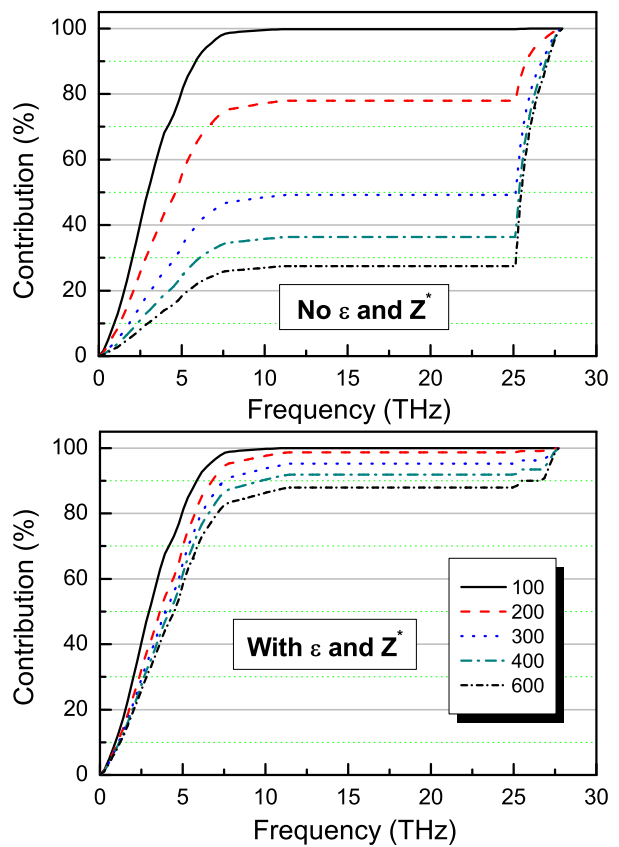}
  \caption{(Color online) At 100, 200, 300,  400 and 600 K, the ratio between  accumulated and  total $\kappa_L$ with respect to frequency  with or without  $\varepsilon$ and $Z^*$.}\label{clc}
\end{figure}

The calculated phonon dispersion of monolayer GeC along high-symmetry path and total and partial density of states (DOS) are shown in \autoref{ph}.
No imaginary frequencies are observed in the phonon dispersion,  indicating  the thermodynamic stability of monolayer GeC.
There are  6 phonon branches due to 2 atoms per unit cell, including
3 acoustic and 3 optical phonon branches. A phonon band gap of 13.52 THz is observed, which separates in-plane transverse optical
(TO) and the in-plane longitudinal optical (LO) branches from out-of-plane optical (ZO), in-plane longitudinal acoustic (LA), in-plane
transverse acoustic (TA) and out-of-plane acoustic (ZA) branches. It is noted that the phonon band gap is larger than width of  acoustic branches (11.61 THz), which has important effects on phonon transport. The large gap  can be explained by the Ge atom  being much heavier
than C atom. It is clearly seen that the ZO branch crosses with the TA and LA branches, which has significant
effect on the phonon  scattering process. The similar phonon dispersion  can also be found  in SiC, ZnO and GaN monolayers\cite{q21,q22,q22-1,q22-2}.
However, for monolayer SiC,  a phonon band gap of 7.47 THz is very smaller than width of  acoustic branches (19.48 THz).
Based on the elastic theory, the  ZA phonon branch should have quadratic dispersion with  the sheet being free of stress \cite{r1,r2}. The quadratic ZA branch near the $\Gamma$ point can be observed for monolayer GeC. However, the TA and LA branches are linear near the $\Gamma$ point.
The partial DOS indicates that optical  branches
are mainly contributed by the vibrations of
C atoms,  while acoustic  branches are  contributed by the
vibrations of Ge atoms.

By solving the linearized phonon Boltzmann equation within single-mode RTA method, the intrinsic $\kappa_L$ of monolayer GeC  is calculated, which is plotted in \autoref{kl} as a function of temperature. The room-temperature $\kappa_L$ of  monolayer GeC is 15.43  $\mathrm{W m^{-1} K^{-1}}$ with the thickness of 18 $\mathrm{{\AA}}$, and the corresponding thermal sheet conductance is 277.74 $\mathrm{W K^{-1}}$,  being two orders of magnitude lower
than that of graphene (about 12884 $\mathrm{W K^{-1}}$)\cite{2dl}. To understand the mechanism underlying the low $\kappa_L$ of
monolayer GeC, the mode level Gr$\mathrm{\ddot{u}}$neisen parameters  of  monolayer GeC are show in \autoref{gru}. It is found that the $\gamma$
of TO and LO branches is fully positive.
For the phonon modes below the gap,  both negative and
partial positive $\gamma$ can be observed, and ZA branch shows very large negative $\gamma$. However, due to the symmetry-based selection rule, the scattering of ZA branche is largely suppressed.
The large  $\gamma$ means strong  phonon anharmonicity, which can produce the strong phonon-phonon scattering, leading to the low $\kappa_L$ of monolayer GeC.
To have an explicit view on the phonon anharmonicity from
another aspect, the potential energy well of monolayer GeC,  defined as  potential energy
change due to the change of lattice constant ($\Delta E$=$E_{a/a_0}$-$E_1$,  where $E_{a/a_0}$ and $E_1$ are the total energies of
strained and unstrained systems, respectively), is shown in  the inset of \autoref{gru}.
It is found that the potential well of monolayer GeC is asymmetric with respect to
 compressive  and tensile strains, which is a direct
evidence of the phonon anharmonicity. A three order polynomial curve can be used to fit the potential energy well, and the fitted parameter for the cubic term is -395 eV, and the large cubic term  is consistent with the $\gamma$.

\begin{figure}
  \includegraphics[width=8cm]{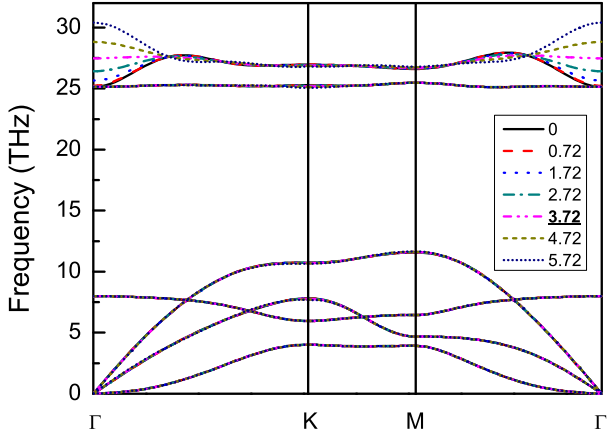}
  \caption{(Color online) Phonon band structures of monolayer GeC with $|Z^*|$ along $xx$ and $yy$ directions artificially changing from 0 to 5.72.}\label{ph1}
\end{figure}

\begin{figure}
  \includegraphics[width=8.0cm]{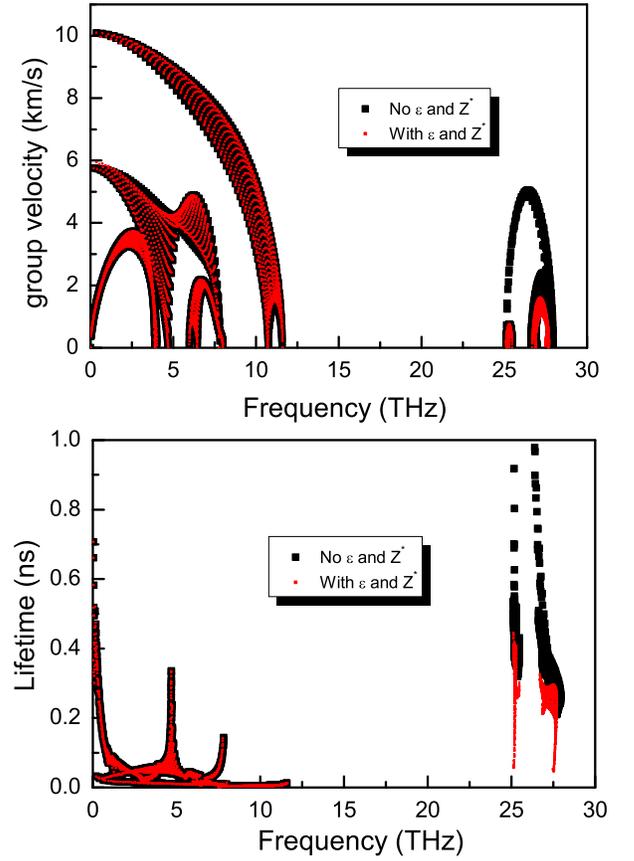}
  \caption{(Color online) The mode level phonon group velocities and phonon lifetimes (300K)  of  monolayer GeC in the first BZ with or without  $\varepsilon$ and $Z^*$.}\label{nac}
\end{figure}

For the most bulk and 2D materials, the  temperature-dependent $\kappa_L$
follows  $\kappa_L$$\sim$$1/T^\alpha$ relationship with  $\alpha$ changing from 0.85 to 1.05\cite{q21}.
It is very strange  that the $\kappa_L$ of monolayer GeC shows an anomalous
linear temperature dependence  above about  350 K, which is obviously different from usual picture $\kappa_L$$\sim$$1/T$ relationship.
The $\kappa_L$$\sim$$1/T$ relation is also shown in \autoref{kl} for comparison.  To understand  the anomalous temperature dependence of $\kappa_L$
of monolayer GeC, the ratio between  accumulated and  total $\kappa_L$ with respect to frequency  at 100, 200, 300, 400 and 600 K are plotted
in \autoref{clc}.  It is clearly seen that, with the temperature increasing, the contribution from LO and TO branches increases.  When the temperatures are lower than 300 K, the contribution of  phonon modes below the phonon gap  is larger than 50\%. However, for
temperatures higher than 300 K, the LO and TO branches dominate the $\kappa_L$ with the contribution larger than 50\%.
If acoustic phonon branches dominate $\kappa_L$, the  temperature-dependent $\kappa_L$ would
follow  $\kappa_L$$\sim$$1/T$ relationship, which can be found  for most materials\cite{q21}. So, the anomalous linear temperature dependence of $\kappa_L$ of monolayer GeC  is  due to dominant contribution from LO and TO branches, when the temperature is larger than 300 K.
Anomalously temperature-dependent $\kappa_L$ has been predicted in monolayer ZnO and GaN by Hu et al.\cite{q21}. To achieve the linear temperature dependence of $\kappa_L$, Hu et al. propose two conditions: (1) With increased temperature, the contribution to $\kappa_L$ from optical
phonon branches should increases.  (2) At high temperature,  the optical branches should dominate $\kappa_L$. The monolayer GeC  indeed satisfies these conditions.

\begin{figure}
  \includegraphics[width=8.0cm]{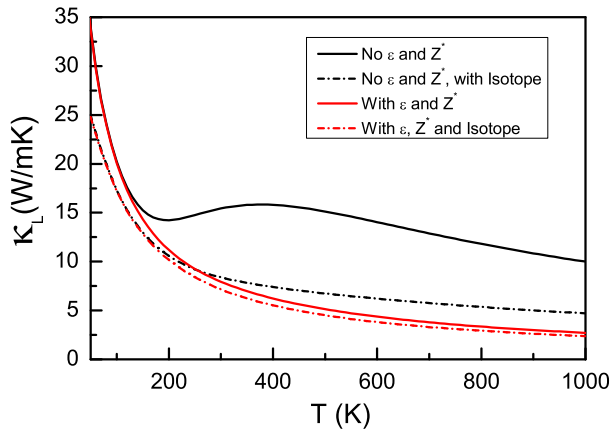}
  \caption{(Color online)  The $\kappa_L$ of  monolayer GeC  with or without  $\varepsilon$,  $Z^*$  and isotope as a function of temperature.}\label{kl-c}
\end{figure}

Due to the large  charge transfer from Ge to C atom, the long-range electrostatic
Coulomb interactions may have important effects on phonon transports of  monolayer GeC.
The large charge transfer can induce strongly polarized covalent bond,  which is also described  by   $Z^*$ and
  $\varepsilon$, as given in \autoref{tab1}. The large $Z^*$ and $\varepsilon$ can produce large  LO-TO splitting (2.35 THz)  at the BZ center, as seen in \autoref{ph}.  The $\kappa_L$ with $Z^*$ and $\varepsilon$ as a function of temperature is also plotted in \autoref{kl}.
It is clearly seen that the relationship between $\kappa_L$ and $T$ changes from  $\kappa_L$$\sim$$T$ to  $\kappa_L$$\sim$$1/T$.
To understand  the sudden change, the ratio between  accumulated and  total $\kappa_L$ with respect to frequency  with $Z^*$ and $\varepsilon$   are also plotted in \autoref{clc} at 100, 200, 300, 400 and 600 K. With increased temperature, the contribution to $\kappa_L$ from high frequency optical
phonon branches increases, but acoustic branches always dominate the $\kappa_L$ for all temperatures ($>$80\%). With inclusion of $Z^*$ and $\varepsilon$,
the condition (2) is broken, so the $\kappa_L$$\sim$$T$ relationship disappears.

To have a better understanding about $Z^*$  effects on $\kappa_L$,
we artificially change   $|Z^*|$ along $xx$ and $yy$ directions  from 0 to 5.72, and the related phonon dispersions and $\kappa_L$ are plotted in \autoref{ph1}
and \autoref{kl}, respectively. The  LO-TO splitting changes from 0 THz to  5.17 THz with increasing $|Z^*|$, and the $\kappa_L$ gradually deviates from the
 $\kappa_L$$\sim$$T$ relationship. When $|Z^*|$ reaches the true values,  $\kappa_L$$\sim$$1/T$ can be observed. To understand the mechanism underlying  the sudden change of $\kappa_L$, the mode level phonon group velocities and phonon lifetimes (300K)  of  monolayer GeC  with or without  $\varepsilon$ and $Z^*$ are plotted in \autoref{nac}. The group velocities and phonon lifetimes of phonon modes below the gap have little difference between with and without  $\varepsilon$ and $Z^*$. However, for LO and TO branches, the group velocities and phonon lifetimes with   $\varepsilon$ and $Z^*$ are smaller than  ones without  $\varepsilon$ and $Z^*$, which dramatically reduces  contribution to $\kappa_L$ from LO and TO branches, and then the normal $\kappa_L$$\sim$$1/T$ relationship is observed in GeC monolayer.

Finally, the phonon-isotope scattering  is consider on $\kappa_L$,  based on the formula proposed by Shin-ichiro Tamura\cite{q24}.  The related $\kappa_L$ with or without $\varepsilon$, $Z^*$  and isotope are plotted in \autoref{kl-c}. Without inclusion of  $\varepsilon$ and $Z^*$, it is clearly seen that the phonon-isotope scattering can also markedly  remove $\kappa_L$$\sim$$T$ relationship. However,  considering $\varepsilon$ and $Z^*$, the phonon-isotope scattering has little effects on $\kappa_L$ of monolayer GeC. In  nanoscale devices,  the  residual strain usually exists, and the substrate  is needed in real applications. Even if the $\varepsilon$ and $Z^*$  are neglected,  the anomalous temperature dependence of $\kappa_L$ should  depend on the strain and specific substrate.
 \begin{figure}
  \includegraphics[width=8.0cm]{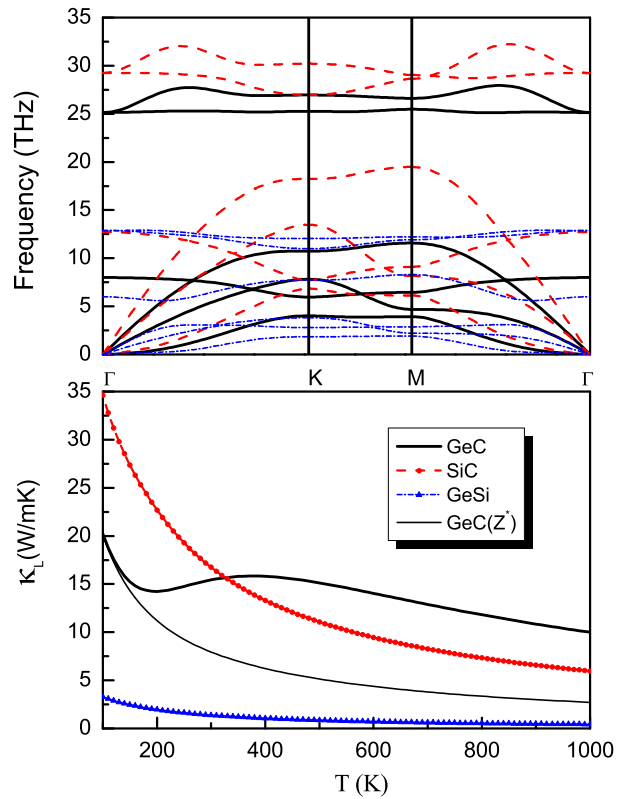}
  \caption{(Color online)Left: phonon band structures of monolayer GeC, SiC and GeSi without  $\varepsilon$ and $Z^*$; Right: The $\kappa_L$ of  monolayer GeC (without and with  $\varepsilon$ and $Z^*$), SiC and GeSi.}\label{c}
\end{figure}

\section{Discussion and Conclusion}
It is interesting to compare phonon transports of monolayer GeC with  ones of SiC and GeSi monolayers. The monolayer SiC has also a perfect planar hexagonal honeycomb structure, but monolayer GeSi has a buckled structure due to the decrease in the overlap between the
$p_z$ orbitals\cite{q22-2}. The phonon band structures and $\kappa_L$ of monolayer GeC, SiC and GeSi without  $\varepsilon$ and $Z^*$ are plotted  in \autoref{c}. The  $\varepsilon$ and $Z^*$ produce little effects on $\kappa_L$ of monolayer SiC and GeSi, so only the $\kappa_L$ of monolayer GeC with  $\varepsilon$ and $Z^*$ is shown in \autoref{c}. It is clearly seen that anomalous temperature dependence of $\kappa_L$ is absent in monolayer SiC and GeSi. These can be understood by their phonon dispersions. There is a  huge phonon band gap of 13.52 THz (larger  than width of  acoustic branches [11.61 THz]) in monolayer GeC, which can  weaken the
scattering between acoustic and high frequency optical phonon modes, producing  very large phonon lifetimes of LO and TO branches.
For monolayer SiC,  a phonon band gap of 7.47 THz is observed (smaller than width of  acoustic branches [19.48 THz]), and the LO and TO modes can be effectively scattered with acoustic modes, leading to short phonon lifetimes of LO and TO branches. In monolayer GeSi, there is a very large gap of 7.17 THz (larger  than width of  acoustic branches [3.81 THz]) between LO/TO and acoustic branches, but ZO branch is in the gap, which provide effective scattering channels for LO and TO branches, giving rise to short phonon lifetimes. When considering $\varepsilon$ and $Z^*$, the order of $\kappa_L$ is GeSi $<$ GeC $<$ SiC, which is consistent with their atomic mass.

In summary,  the phonon transports of monolayer GeC are investigated by   the first-principles calculations and semiclassical Boltzmann transport theory.
When neglecting  $\varepsilon$ and $Z^*$, monolayer
GeC  possesses anomalously linear temperature dependent $\kappa_L$, which is  different from the commonly established $\kappa_L$$\sim$$1/T$ relationship.
The large deviation is because the
contribution to $\kappa_L$  from LO and TO branches increases with increasing  temperature, and eventually dominates the $\kappa_L$ with $T$ being larger than 300 K. However, considering $\varepsilon$ and $Z^*$, the common $\kappa_L$$\sim$$1/T$ relationship is observed by reduced  group velocities and phonon lifetimes of LO and TO branches.  It is found that the  phonon-isotope scattering can also weaken  anomalously linear temperature dependent $\kappa_L$ in monolayer GeC. This work presents a comprehensive understanding of the phonon transports of monolayer GeC, and sheds light on further studies of phonon
transports of other 2D materials.

\begin{acknowledgments}
This work is supported by the National Natural Science Foundation of China (Grant No.11404391).  We are grateful to the Advanced Analysis and Computation Center of CUMT for the award of CPU hours to accomplish this work.
\end{acknowledgments}

\end{document}